# On the impact of capillarity for strength at the nanoscale

Nadiia Mameka[1*], Jürgen Markmann[1,2], and Jörg Weissmüller[1,2*]
*1–Institute of Materials Research, Materials Mechanics,*
*Helmholtz-Zentrum Geesthacht, Geesthacht, Germany* and
*2–Institute of Materials Physics and Technology,*
*Hamburg University of Technology, Hamburg, Germany*[*]

The interior of nanoscale crystals experiences stress that compensates the capillary forces and that can be large, in the order of 1 GPa. Various studies have speculated on whether and how this surface-induced stress affects the stability and plasticity of small crystals. Yet, experiments have so far failed to discriminate between the surface contribution and other, bulk-related size effects. In order to clarify the issue, we study the variation of the flow stress of a nanomaterial while distinctly different variations of the two capillary parameters surface tension and surface stress are imposed under control of an applied electric potential. Our theory qualifies the suggested impact of surface *stress* as not forceful and instead predicts a significant contribution of the surface energy, as measured by the surface *tension*. The predictions for the combined potential- and size dependence of the flow stress are quantitatively supported by the experiment. Previous suggestions, favoring the surface stress as the relevant capillary parameter, are not consistent with the experiment.

## I. INTRODUCTION

Even when there is no external load, the interior of a small crystal experiences a "surface-induced" stress, $\boldsymbol{\sigma}_\mathsf{C}$, which compensates the capillary forces that are quantified by the surface stress, $f$ [1]. With the magnitude of $\sigma_\mathsf{C}$ in the order of $f/r$ [2] and $f$ in the order of 3 N/m [3], stresses in excess of 1 GPa are expected for crystals with characteristic radius, $r$, at the lower nanoscale. Since $\boldsymbol{\sigma}_\mathsf{C}$ typically has a significant deviatoric stress component [2], the capillary forces may affect the shear deformation of crystal plasticity. Indeed, an instability to spontaneous plastic deformation is observed in extremely small structures. Atomistic simulation studies [4–8] and, on a more speculative note, experimental reports [9, 10] attribute the instability to plastic shear prompted by the action of the surface-induced stress. It has also been suggested that $\boldsymbol{\sigma}_\mathsf{C}$ will enhance the action of a compressive external load and diminish that of a tensile load, resulting in a substantial tension-compression asymmetry in the strength of nanowires [6, 11–13] and in the strength [7, 8] as well as creep rate [14] of nanoporous metals. Thus, surface stress is believed to impose a lower limit on the stable size of crystals and to contribute substantially to nanoscale mechanical behavior. Here, we critically examine the suggested impact of the surface stress. We analyse the relevant aspects of the mechanics of small crystals and we link this analysis to experiments in which the surface stress is varied in situ during the deformation of a nanomaterial and the impact of that variation on the flow stress monitored. The results do not support the suggested impact of *surface stress* on strength. Instead, they agree with the predicted action of a separate capillary parameter, the specific excess free energy or *surface tension*, which is not typically considered in this context.

It is known that the surface tension, $\gamma$, may prompt spontaneous shortening of macroscopic metal wires by creep at elevated temperature. Zero creep measurements, pioneered by H. Udin around 1950 [15, 16] and later extended to multilayers [17], measure $\gamma$ via the tensile load required to suppress the contraction. The impact of $\gamma$ on plasticity is further emphasized by studies of engineering materials wetted by electrolytes. These reveal similarities between the electrode-potential dependence of $\gamma$ and creep rate [18, 19] or fracture stress [20]. Zero creep experiments typically use wires a few tens of $\mu$m in diameter and very low stresses, in the order of 10-100 kPa. Yet, as the impact of surface phenomena is enhanced at small size, much larger surface-related stresses may be expected for nanowires.

The net surface excess free energy, $G_\mathsf{S}$, scales with the surface area, $A$. Wires tend to contract spontaneously in zero creep experiments since the contraction reduces $A$ and, thereby, the energy: $\delta G_\mathsf{S} = \gamma \delta A$. The stress which is required for compensating the trend for contraction – resulting in zero creep rate – in a wire of radius $r$ is tensile and of magnitude $\gamma/r$ [15, 16]. Zero creep experiments thus exemplify that tension-compression asymmetry results from the action of capillarity: creep is arrested by tensile stress but would be accelerated by a compressive stress of same magnitude. It is less obvious how the notion of a shear deformation driven by *surface stress* connects to energy minimization. In fact, plastic shear may create slip steps or terraces of new crystallographic orientation, both of which may in principle *increase* $G_\mathsf{S}$, excluding a spontaneous process. Suggestions of surface stress as a driving force for spontaneous shear have so far not been linked to the energetics, and the present work addresses the issue.

Recent studies of the deformation of nanoporous gold (NPG) in situ in electrolyte present new opportunities for investigating nanoscale mechanical behavior by experiment. NPG is an emerging model nanomaterial that can be made with mm dimensions and tested using re-liable macroscopic testing schemes [21]. The polycrystalline material with 10-100 $\mu$m grain size is distinguished by its network structure of nanoscale struts or "ligaments". The brittle failure of NPG in tension relates to fracture mechanics concepts such as the distribution of heterogeneities in the network structure [22]. By contrast, the material's excellent deformability in compression provides opportunities for probing the mechanisms and driving forces of yielding and plastic flow in small scale plasticity. In fact, the mechanical behavior of the ligaments agrees well with that of gold nanopillars and nanowires [21, 23–25], supporting the relevance of studies of NPG for understanding small-scale plasticity in general. In situ tests of NPG in electrolyte allow monitoring the mechanical behavior while the surface state is modulated under control of the electrode potential, $E$ [26–28]. As it is known how the capillary forces $\gamma$ and $f$ vary independently with $E$ [29], such experiments–as described in this study–can unravel the impact of the two capillary parameters.

[*] Authors for correspondence: nadiia.mameka@hzg.de, weissmueller@tuhh.de

## II. THEORY

Our analysis admits that a part, $\Delta T$, of the external applied stress, $T$, that drives the deformation is required for doing work against as-yet unspecified capillary forces. This part is not available for overcoming the intrinsic dissipative forces that determine the resistance to dislocation motion in bulk plasticity. The apparent experimental flow stress, $\sigma^{\text{flow}}$, is then the sum of the intrinsic bulk flow stress, $\sigma_0^{\text{flow}}$, and of $\Delta T$:

$$\sigma^{\text{flow}} = \sigma_0^{\text{flow}} + \Delta T. \quad (1)$$

We now explore two possible origins of $\Delta T$.

We analyze a solid with volume $V$ and surface area $A$, which are both measured in stress-free states of the solid. The total free energy may be expressed as $G = V\Psi + A\gamma$ with $\Psi$ the volumetric free energy density in the bulk. As the defining equation for $\gamma$ we may thus take

$$\gamma = G_{\text{S}}/A = (G - V\Psi)/A, \quad (2)$$

so that $\gamma$ is the excess, per area, in free energy over that of a bulk solid with same volume but negligible surface effects. If the plastic strain $\delta\varepsilon^{\text{p}}$ changes $A$ or $\gamma$, then a part, $\Delta T V \delta\varepsilon^{\text{p}}$, of the mechanical work is consumed for supplying the extra energy $\delta(\gamma A)$. Equating mechanical work and free energy change yields the required extra traction as

$$\Delta T = \frac{1}{V}\frac{\delta G_{\text{S}}}{\delta\varepsilon^{\text{p}}} = \gamma\frac{\delta\alpha}{\delta\varepsilon^{\text{p}}} + \alpha\frac{\delta\gamma}{\delta\varepsilon^{\text{p}}} \quad (3)$$

with $\alpha = A/V$ the volume-specific surface area. Contrary to the dissipative processes of classic plasticity, the impact of surface tension on the flow stress links to a conservative process that stores or releases energy.

The capillary parameter that relates to elasticity is the surface stress, $f$. It quantifies the tendency of the surface to compress ($f > 0$) or expand ($f < 0$) the solid elastically. Restricting attention to isotropic surfaces we take $f = d\gamma/de$ with $e$ the relative change in surface area (in laboratory coordinates) by tangential elastic strain.

As a model which incorporates the most obvious features of a small-scale solid we consider the long (negligible end-effects) cylindrical nanowire of radius $r$ and length $l \gg r$, for which $\alpha = 2/r$. We take elasticity, surface tension and surface stress as isotropic.

Even in the absence of an applied load, the surface stress requires a compensating stress $\boldsymbol{\sigma}_{\text{C}}$ in the bulk of the nanowire; its axial and radial components are [2] [30]

$$\sigma_{\text{C}}^{\text{A}} = -\alpha f \qquad \text{and} \qquad \sigma_{\text{C}}^{\text{R}} = -\tfrac{1}{2}\alpha f, \quad (4)$$

respectively. They prompt the surface-induced elastic strain $\boldsymbol{\epsilon}_{\text{C}}$, with axial and radial components [31]

$$\epsilon_{\text{C}}^{\text{A}} = \frac{\nu-1}{Y}\alpha f \qquad \text{and} \qquad \epsilon_{\text{C}}^{\text{R}} = \frac{3\nu-1}{2Y}\alpha f. \quad (5)$$

$Y$ and $\nu$ represent Young's modulus and Poisson's ratio, respectively, of the bulk. Consequences of this surface-induced elastic relaxation are $i$) a reduction in the energy of the surface regions by $f\delta e$, where $\delta e = \epsilon_{\text{C}}^{\text{A}} + \epsilon_{\text{C}}^{\text{R}}$ and $ii$) an increase of the elastic strain energy density in the bulk by $\delta\Psi = \tfrac{1}{2}\boldsymbol{\epsilon}_{\text{C}} : \boldsymbol{\sigma}_{\text{C}}$. It is well known [32, 33] that the energy increase in the bulk can only partly compensate the energy reduction at the surface. Inserting the two energy terms into Eq 2, one indeed finds a reduced $\gamma$ of the relaxed nanowire,

$$\gamma_{\text{relaxed}} = \gamma_0 - \frac{3-5\nu}{4Y}f^2\alpha. \quad (6)$$

where $\gamma_0$ refers to the unstrained surface.

The contribution of capillarity to the flow stress of a nanowire, loaded axially, is readily obtained by evaluating Eq 3 while using Eq 6 for $\gamma$ and noting that $\delta\alpha/\delta\varepsilon^{\text{p}} = \alpha/2$ for a long cylinder (elongation at constant $V$ increases the surface area). One thus obtains

$$\Delta T = \frac{1}{2}\alpha\left(\gamma_0 - \frac{3-5\nu}{2Y}f^2\alpha\right). \quad (7)$$

The relaxation terms in Eqs 6 and 7 are small in any case and negligible in experimental situations even for very small structures [34], suggesting that surface stress does not contribute significantly to $\Delta T$. Ignoring the $f$-dependent term in Eq 7 we find that $\Delta T_{\text{S}}$, the change in flow stress due to the surface excess free energy, is simply

$$\Delta T_{\text{S}} = \frac{\gamma_0}{r}. \quad (8)$$

This is the well-known relation behind zero-creep experiments. In view of Eq 1 and of the sign convention, positive and negative stress in tension and compression, respectively, Eq 8 suggests strengthening in tension yet weakening in compression, in other words, a tension-compression asymmetry of the contribution of the surface to stresses in small-scale plasticity.

Note that Eq 8 accounts for an *energy* balance during plastic deformation and is not inherently related to the acting *stresses* in the nanowire. Those stresses are given by Eqs 4, and they scale with $f$ rather than $\gamma$.

Even though the energy-based considerations marginalize the role of surface stress, let us inspect a conceivable direct impact of that parameter on the deformation of a nanowire. Of relevance for plasticity is only the deviatoric part of $\boldsymbol{\sigma}_\mathsf{C}$. Equations 4 imply this to be a uniaxial stress, of magnitude $-\alpha f/2$, along the wire axis. This stress adds to that caused by the external traction. The Peach - Köhler forces on dislocations in the interior of the wire then see an extra contribution, analogous to the addition of an external stress $\Delta T_\mathsf{C}$, which obeys

$$\Delta T_\mathsf{C} = \frac{f}{r}. \qquad (9)$$

As explained in the Introduction, the reasoning behind Eq 9 has given rise to suggestions – partly by one of the present authors – that the surface stress enhances the tensile strength of nanowires or may even prompt spontaneous plastic contraction [4–9, 11–13]. Yet, this argument is problematic since it singles out the action of the bulk stress on the dislocations, thereby ignoring the action of the stresses near the surface.

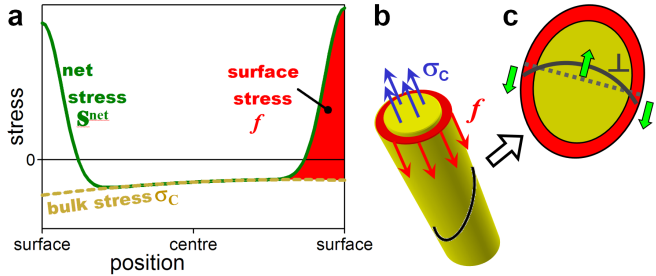

FIG. 1. Stress in a nanowire and its impact on dislocation glide (schematic). **a**, Green line, stress profile along a linear section through the wire. The actual stress $\mathbf{S}^\mathrm{net}$ may be decomposed into bulk stress $\boldsymbol{\sigma}_\mathsf{C}$ (yellow dotted line) and surface stress $f$ (red shaded area). **b**, Balance of force on a normal cross-section. Surface regions (red) experience tensile stress, which is represented by the surface stress and which is compensated by an oppositely-signed surface-induced stress $\boldsymbol{\sigma}_\mathsf{C}$ in the bulk (blue). **c**, Dislocation (grey line) on an inclined cross-section. Shear components of stresses from (a) give rise to Peach-Köhler forces which mutually compensate.

Figures 1a) and b) illustrate how the continuum theory of capillarity decomposes the position-dependent net stress, $\mathbf{S}^\mathrm{net}$, in a nanowire into the bulk stress $\boldsymbol{\sigma}_\mathsf{C}$ that acts throughout the cross-section and surface stresses that act along its perimeter. Let us here ignore this decomposition and relate the energetics of dislocation plasticity to the more fundamental quantity $\mathbf{S}^\mathrm{net}$. In the absence of an external load, mechanical equilibrium requires that the area-integral of the traction, $\mathbf{t} = \mathbf{S}^\mathrm{net} \cdot \mathbf{n}$, on a cross-section (unit normal $\mathbf{n}$) through the nanowire must vanish. One can readily confirm that the net mechanical work which is done by the Peach-Köhler forces when a dislocation glides over the entire cross-section scales with the integrated traction force [35] and so must vanish when there is no external load. Figure 1c) illustrates the opposite-signed Peach-Köhler forces on dislocation segments in the bulk and near the surface, as implied by the opposite-signed stresses in the respective regions. These forces act analogously on full dislocations and on partial dislocations that propagate a stacking fault or a twin. The lattice instability of small nanowires by twinning shears the entire cross-section by a partial dislocation Burgers vector. The work against the acting stresses is again governed by the area integral of the traction [36], which vanishes.

A quite different result and specifically a nonvanishing work of deformation would be obtained if, erroneously, the net mechanical work was derived from the bulk stress $\boldsymbol{\sigma}_\mathsf{C}$ alone, excluding the stress in the surface regions from the consideration. This, however is the argument that leads to the prediction of a net contribution of the surface-induced bulk stress to dislocation plasticity, Eq 9, and to the suggestion of a spontaneous plastic shear driven by surface stress. Clearly, that approach is not appropriate, and claims of a surface-stress induced strengthening or weakening of nanowires must be considered with caution or even rejected outright.

While dislocations from a stable or increasing population, for instance sustained from single-arm sources [37], may carry the plasticity and control the strength of small structures including NPG [8, 38–40], nanowires may be dislocation-starved and their strength controlled by dislocation nucleation [6, 41]. As nucleation is favoured *at free surfaces* of bulk materials [36] and nanowires [6, 41], the nucleation events do not probe the surface-induced bulk stress that leads to Eq 9 but they are at least partly affected by the large and opposite-signed stresses in the surface regions. This again sheds doubt on predictions, such as Eq 9, for strengthening or weakening by surface stress, emphasizing the need for experiment.

Discriminating by experiment between the impact of surface stress and surface tension on the plastic flow of nanostructures is challenging since Eqs 8 and 9 predict typically quite similar size-effects. However, one may exploit that $\gamma$ and $f$ respond differently to changes in the electrode potential, $E$, or in its conjugate parameter, the superficial electric charge density (charge per area) $q$, if the surface is wetted by a fluid electrolyte. For a recent review of this "electrocapillary coupling" see Ref. [29].

The Lippmann equation requires that $\mathrm{d}\gamma = -q\mathrm{d}E$. In as much as the capacitance, $c$, can be approximated as constant near the potential of zero charge, $E_\mathrm{zc}$ (where

$q = 0$), $\gamma$ varies parabolically as

$$\gamma = \gamma_{zc} - \frac{1}{2} c \left( E - E_{zc} \right)^2 . \qquad (10)$$

Contrary to $\gamma$, the surface stress may vary linearly near $E_{zc}$, so that (see Eq 5.34 in Ref [29])

$$f = f_{zc} + \varsigma c \left( E - E_{zc} \right) , \qquad (11)$$

with $\varsigma$ the electrocapillary coupling coefficient. The values of $c$ and $\varsigma$ are well established for gold surfaces in weekly adsorbing aqueous electrolytes, such as those of the present work, near $E_{zc}$. Here $c \approx 40 \mu F/cm^2$ [42]; furthermore, $\varsigma$ is invariably negative valued at transition metal surfaces near $E_{zc}$, and specifically $\varsigma = -2.0$ V for gold [3, 29, 43, 44]. Figure 2a compares the variation of $\gamma$ and of $f$ for gold surfaces near $E_{zc}$.

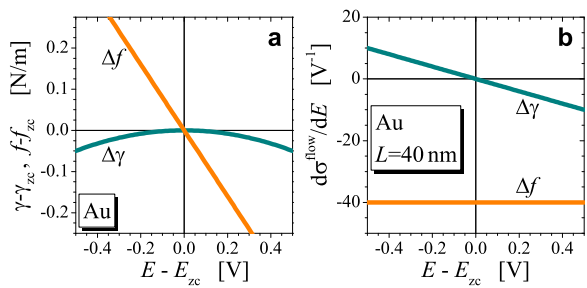

FIG. 2. Capillary forces at gold surfaces and their impact on the flow stress of a nanowire. **a**, Linear variation of surface stress, $f$, with electrode potential, $E$, around the potential of zero charge, $E_{zc}$, is distinguished from parabolic variation of the surface tension, $\gamma$. **b**, Flow-stress potential coupling parameter, $d\sigma^{flow}/dE$, versus $E$. Qualitatively different predictions for the coupling are obtained depending on whether $\sigma^{flow}$ is assumed to respond to changes in $\gamma$, Eq 12, or in $f$. Graphs show extrapolated behavior based on the quadratic and linear approximations of Eqs 10 and 11 along with experimental values for capacitance and electrocapillary coupling of gold near $E_{zc}$, see main text. Part b) assumes ligament diameter 40nm, representative of the experiment.

The experiments in this work explore the variation of the flow stress, $\sigma^{flow}$, with $E$. Figure 2b summarizes the implications of our discussion for the coupling $d\sigma^{flow}/dE$, accounting for the numerical values of $c$ and of $\varsigma$ of gold near $E_{zc}$. If the surfaces affect the strength via surface stress, then Eqs 9 and 11 imply $d\sigma^{flow}/dE = c\varsigma/r$. As $c > 0$ and $\varsigma < 0$ on clean transition metal surfaces, it follows that $d\sigma^{flow}/dE$ is *negative* throughout the potential regime of capacitive charging. By contrast, if surface tension is the relevant capillary parameter (Eq 8), then Eq 10 implies that

$$\frac{d\sigma^{flow}}{dE} = -\frac{c}{r} \left( E - E_{zc} \right) . \qquad (12)$$

Here, the stress-potential coupling is *positive* at potentials negative of $E_{zc}$, yet the sign is inverted when $E_{zc}$

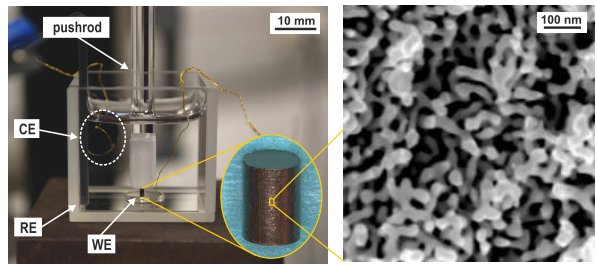

FIG. 3. In situ setup for compression tests under potential control. **Left**, Nanoporous gold (NPG) sample (inset) forms the working electrode (WE) and is loaded by a glass rod. CE and RE: counter and reference electrodes. **Right**, Scanning electron micrograph of the NPG microstructure.

is crossed. The distinctly different predictions will allow us to discriminate, by means of in situ deformation experiments in electrolyte, between the two scenarios.

## III. EXPERIMENT

As detailed in the Methods Section, we prepared macroscopic samples of NPG with different mean ligament diameters, $L$, and with solid fractions $\varphi \sim 0.3$ by electrochemical dealloying. The setup of Fig 3 allowed uniaxial compression tests in situ in electrolyte and under control of the electrode potential, $E$. Motivated by the distinctly different behavior of surface stress and surface tension during *capacitive* charging, see Section II, we focused on potentials in the vicinity of $E_{zc}$. Our electrolytes, 0.7 M NaF, 1 M HClO$_4$ and 0.5 M H$_2$SO$_4$, comprise anions that adsorb nonspecifically on Au. Yet, the strengths of the gold-anion interactions differ [3]. Potential steps were imposed during compression. Cyclic voltammograms (see Fig S1 in the Supporting Online Material, SOM) show the region of dominantly capacitive charging to extend up to $E \leq 1.0$ V, while electrosorption of OH$^-$, involving up to one molecular monolayer [45], dominates at more positive $E$. All electrode potentials in this work are referred to the standard hydrogen electrode (SHE).

Figure 4 summarizes exemplary results of an in situ compression test, here for $L = 40$ nm and in 0.5 M H$_2$SO$_4$. The potential (blue line in Fig 4a) was kept at 1.0 V up to 20% engineering strain, establishing a reference for deformation at constant potential. $E$ was then stepped to its most negative value, 0 V, and a series of potential holds, separated by 100 mV, followed. The step sequence was inverted when $E$ reached 1.5 V. The stress-strain graph in Fig 4a illustrates how the experimental (flow-) stress, $\sigma$, reacts to the variation of $E$. The large deformability and pronounced strain hardening agrees with findings for compression of NPG in air [21]. The most obvious consequence of the potential variation is the strong change in $\sigma$ during oxygen electrosorption (shaded regions in Fig 4), during the jumps at low potential (Fig 4b) but it is enhanced at higher potential (Fig 4c). This behavior is more obvious when inspecting the flow-stress potential response parameter, $\delta\sigma/\delta E$, in Fig 4d. In the capacitive regime $\delta\sigma/\delta E$ starts out positive-valued at negative $E$. Increasing $E$ lets the response approach zero and then change sign at around 0.7 V.

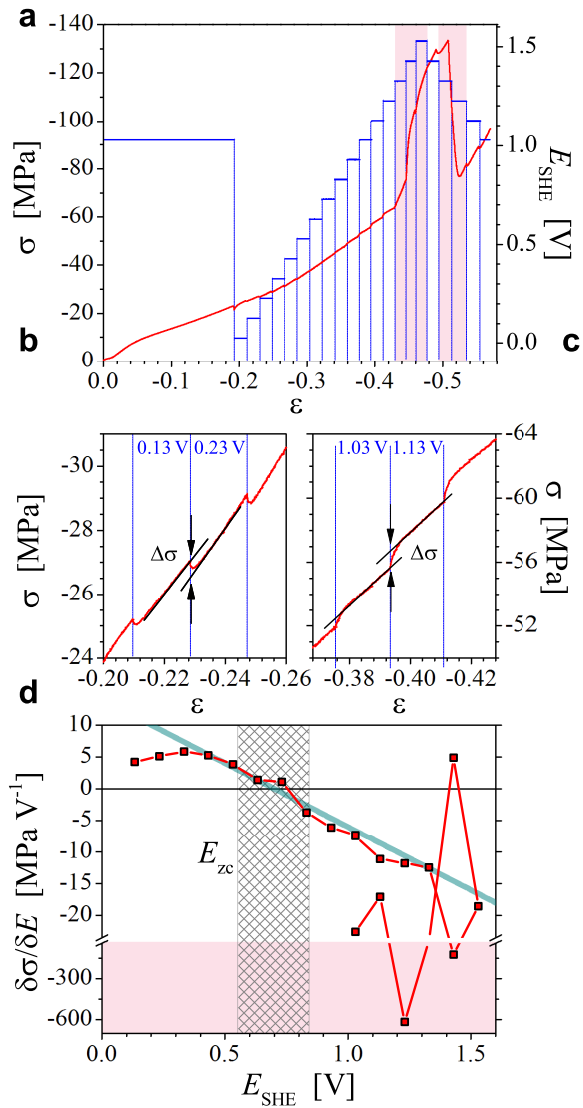

The sign change of $\delta\sigma/\delta E$ is remarkable in view of the inversion of the potential-response of the surface tension at the potential of zero charge, $E_{zc}$, see Eq 10. The value of $E_{zc}$ is characteristic of the combination of electrolyte and of the electrode surface's crystallography and defect structure. Using two independent variants of the immersion technique (see Methods and Fig 6b below), $E_{zc}$ of NPG in 0.5 M $H_2SO_4$ was determined as $0.70 \pm 0.14$ V. This value indeed coincides with the potential where $\delta\sigma/\delta E$ inverts its sign, see Fig 4d.

The bold solid line in Fig 4d represents the prediction of Eq 12 for $\delta\sigma^{flow}/\delta E$ of gold nanowires with $L = 40$nm, accepting $\gamma$ and not $f$ as the governing capillary parameter and using $c = 40\mu F/cm^2$ (see Section II). It is striking that, around $E_{zc}$, the slope of the experimental graph is in excellent agreement with the prediction.

Figure 5 summarizes the results of in situ compression tests with different $L$. Strength and flow stress increase well compatible with the observations in Ref. [26]. Yet, the focus of the present study is on the capacitive processes. The jumps in stress are here smaller, but the enlarged graphs of Fig 4b and 4c show that they are well detectable.

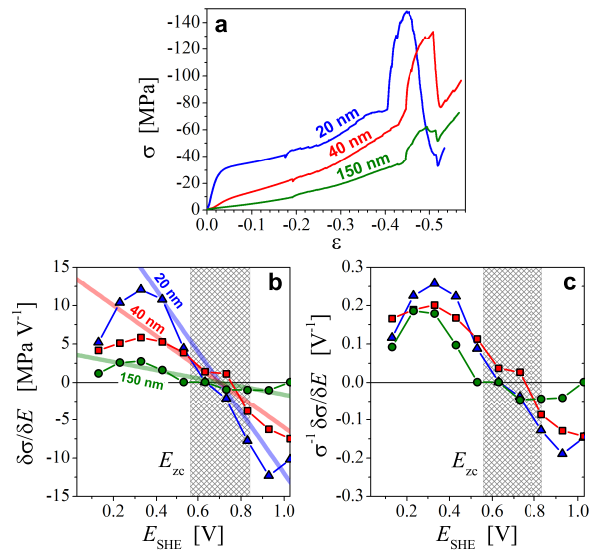

FIG. 4. Results of an in situ compression test in 0.5 M $H_2SO_4$. **a**, Red, stress $\sigma$ versus strain $\varepsilon$ at constant engineering strain rate $10^{-5}$/s. Potential, $E$, vs. SHE (blue) is superimposed. **b,c**, Details from (a), showing increase or decrease of $\sigma$ in response to potential steps during capacitive charging. **d**, Response, parameterized as $\delta\sigma/\delta E$, of experimental flow stress to jumps in $E$. Note axis break and different scales for the capacitive- and OH-regions. Blue solid line: predicted response, Eq 12; note excellent agreement near potential of zero charge, $E_{zc}$. Hatched: experimental range of $E_{zc}$. Red shaded regions in **a** and **d** denote regimes of OH-adsorption/desorption. Mean ligament diameter is 40 nm.

FIG. 5. Experiments with different ligament size, as indicated by labels, in 0.5 M $H_2SO_4$: **a**, Flow stress $\sigma$ vs. strain $\varepsilon$ during deformation with strain rate $10^{-5}$/s in compression. **b**, Effective response of $\sigma$ to jumps in the electrode potential, $E$, determined as $\delta\sigma/\delta E$. Straight lines: prediction by Eq 12. **c**, The parameter $\delta\sigma/\delta E$ is normalized to the actual value of flow stress, $\sigma_0$. Note no size-dependent behavior in this case. Shaded: range of potential of zero charge, $E_{zc}$.

Remarkably, the flow stress magnitude is diminished with decreasing $L$, in agreement with previous reports [21, 23–25]. We again focus on $\delta\sigma/\delta E$ during capacitive charging, see Fig 5b. The general trends agree well for all $L$, yet the response is stronger for smaller $L$. The size dependence is anticipated by Eq 12, and indeed the solid lines which represent that equation in Fig 5b (no free parameters) agree quantitatively with the experiment.

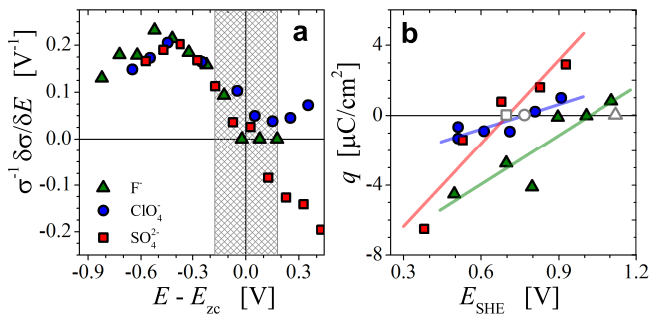

FIG. 6. Experiments with different anions, as indicated in legend, for samples with mean ligament diameter 40 nm. **a**, Normalized flow-stress electrode-potential coupling parameter $\sigma^{-1}\delta/\delta E$, plotted versus difference, $E - E_{zc}$, between electrode potential and the potential of zero charge, $E_{zc}$. Note the excellent agreement in the capacitive regime. **b**, Immersion charge density, $q$, versus electrode potential. Intersects of linear fits with the abscissa provide $E_{zc}$. Open symbols: $E_{zc}$ from separate experiments using open-circuit immersion, see Methods. Shaded: error range for $E_{zc}$.

Fig 5b also shows that $\delta\sigma/\delta E$ exhibits the same size-dependent trend as $\sigma$. In order to verify this observation, Fig 5c plots the response parameter normalized to the flow stress, $\sigma^{-1}\delta\sigma/\delta E$, versus $E$. The graphs nearly coincide in this representation, irrespective of $L$. Apparently, there is a link between phenomena responsible for strengthening by electric potentials and those governing the size-dependence of the strength.

We also explored two other electrolytes, aqueous 1 M HClO$_4$ and 0.7 M NaF. Figure 6a shows the corresponding normalized response parameters $\sigma^{-1}\delta/\delta E$, along with the result for H$_2$SO$_4$. The ligament size was 40 nm for each sample. In order to remove the impact of the different $E_{zc}$ in the individual electrolytes, all electrode potentials were referred to the respective $E_{zc}$. The $E_{zc}$ for our studies with SO$_4^{2-}$, ClO$_4^-$, and F$^-$, as determined by the immersion method (see Fig 6b and Methods) were $0.70 \pm 0.14$, $0.77 \pm 0.19$, and $1.02 \pm 0.22$ V. Figure 6a shows that highly reproducible results, independent of the anion, are obtained when $\sigma^{-1}\delta\sigma/\delta E$ is plotted versus $E - E_{zc}$. Specifically, the trend for the response to change sign at $E_{zc}$ appears generic.

Results of additional in situ compression tests, Fig S2 in the SOM, confirm that the sign-inversion of the flow-stress potential response is recovered when scanning twice through $E_{zc}$ and that at all strains (down to values as small as 4%) the sign of the response is consistent with Eq 12 and with the prediction of the "$\Delta\gamma$" graph of Fig 2. Thus, all experiments support surface tension as the relevant capillary force. The stronger and oppositely-signed response that would indicate surface stress as relevant (see the "$\Delta f$"-graph in Fig 2) is not supported by the experiment.

Our theory presupposes that plastic deformation changes the net surface area, $A$. In order to verify this notion we used the electrochemical capacitance ratio method (see SOM) to monitor the evolution of $A$ during compression experiments such as Fig 4 and 5. The results, as described by Fig S3a in the SOM, confirm that $A$ diminishes continuously during compression. The relative change in $A$ (Fig S3b) during compression is in fact consistent with the atomistic simulation of Ref [8]. Electron micrographs (Fig S4) rule out potential-induced coarsening as an origin of the variation. Thus, besides densifying the ligament network [8, 38], the plastic compression changes the microstructure by reducing the ligament aspect ratio, which decreases the net surface area. Furthermore, previous experiment [38, 39] and atomistic [8] as well as continuum simulation [40] suggest that compression also enhances the dislocation density.

## IV. DISCUSSION

**Apposing experiment and theory.** Contrary to suggestions in previous work, our theory finds no forceful argument for a significant impact of surface stress on the plastic flow of nanowires. Our arguments rest on *i)*, an explicit consideration of the local stress state of the material near the surface and of its impact on dislocation plasticity and *ii)*, the analysis of the energy of the deformed nanowire, in which the contributions of the surface tension dominate while contributions due to surface-stress induced relaxation are negligible. In as much as surfaces contribute a driving force for plastic deformation, the more obvious key parameter is the surface tension $\gamma$ and not the surface stress $f$.

By deforming NPG wetted by electrolyte under potential control, our experiments probe the flow stress variation while $\gamma$ and $f$ can be varied simultaneously yet in distinctly different manner. NPG, when unloaded at any state of plastic flow and then reloaded, yields at the last flow stress before the unload (see Ref [21] and Fig S5 in the SOM). Thus, the flow stress at any state of plastic strain agrees with the yield stress in that state; this connects our experimental investigations of plastic flow to the strength of nanostructures.

The experiment supports our theory: *i)*, for negatively charged surfaces, our experiment finds a positive-valued coupling, $d\sigma^{\text{flow}}/dE$, between flow-stress and electrode potential. The sign agrees with the prediction based on $\gamma$ as the relevant capillary parameter but is incompatible with the potential-dependence of $f$ (see Fig 2). *ii)*, the sign inversion of $d\sigma^{\text{flow}}/dE$ at the potential of zero charge, $E_{zc}$, in our experiment is consistent with the variation of $\gamma$ as embodied in the Lippmann equation and it disagrees with the expectation that $d\sigma^{\text{flow}}/dE$ should be negative-signed at all potentials if $f$ were the controlling quantity. In fact, *iii)*, our in situ capacitance data show the surface area to decrease during plastic deformation, in agreement with the premises behind the analysis of $\gamma$ as a driving force for plastic compression. The decrease of the surface area of NPG with strain is least pronounced in the early stages of deformation, see Ref [8] and Fig S3. Consistent with our theory, the flow stress potential response also tends to be less pronounced at small strain (see the data at most negative potential in Figs 5c, 6a, and S2 d). *iv)*, as the most compelling evidence, the potential- and size- dependence of the experimental coupling strength near $E_{zc}$ agree quantitatively with the prediction of Eq 12.

**Potential of zero charge.** Determining potentials of zero charge is notoriously challenging and this motivates a critical inspection of our findings. $E_{zc}$ of single- and polycrystalline gold surfaces in similar electrolytes tend be 200 to 500 mV more negative than our data. For instance, capacitance measurements in 0.01 M $HClO_4$ suggest 470 and 320 mV for bulk-truncated Au(111) and Au(100), respectively [46]. Immersion measurements of $E_{zc}$ for polycrystalline gold in dilute $HClO_4$, $H_2SO_4$ or NaF solutions have been found even more negative, from 170 to 330 mV [47]. The more positive values of our study may suggest that the extremely high defect density (step edges, kinks)–which is required by the curvature of the surfaces of NPG–shifts $E_{zc}$ to positive. Our data is also qualitatively consistent with the expected compressive strain (Eq 5) in NPG and with the negative-valued $\varsigma$. Furthermore, the order of the $E_{zc}$ for the different anions ($SO_4^{2-} \lesssim ClO_4^- < F^-$) agrees well with literature data [48–50]. Lastly, in agreement with theory, the results for $d\sigma^{\text{flow}}/dE$ in experiments with the individual ion species coincide precisely when plotted versus $E-E_{zc}$, see Fig 6a). Thus, even when the uncertainties involved in determining $E_{zc}$ for real surfaces of the nanoporous metals in our compression experiments are acknowledged, the general magnitude of our $E_{zc}$ and the variation with the anion species appear robust.

The above arguments support the notion that the minimum of the flow stress magnitude (at constant strain rate) in our compression experiments is connected to $E_{zc}$. This is compatible with early experiments on macroscopic metal wires, which exposed a minimum of the tensile creep rate (at constant stress) at $E_{zc}$ [18, 51]. Both findings are indeed consistent with the theory, if the predicted tension-compression asymmetry is born in mind.

**Conservative versus dissipative processes.** The analysis of *conservative* (energy-related) processes during plastic flow leads to the prediction for the change in flow stress due to the surface excess free energy (or surface tension, $\gamma$), Eq 12. The predicted change is positive, suggesting a tension-compression asymmetry with strengthening in tension and weakening in compression. The well-established finding that smaller is stronger in both, tension and compression, implies that additional, *dissipative* strengthening processes act in a symmetric manner. The nature of the dissipative processes is not the subject of our work. However, it is significant that we find the potential-response of the flow stress to scale with the flow stress itself, even for different structure size where values of $\sigma^{\text{flow}}$ differ significantly because smaller is stronger. This scaling at least rules in that the dissipative contributions to $\sigma^{\text{flow}}$ are also related to the surface, a scenario that would result in a size-dependent $\sigma^{\text{flow}}$. In fact, the scenario is compatible with previous observations from *in situ* tests in electrolyte studying the impact of the specific adsorption of $OH^-$ ions. These experiments where rationalized [26] in terms of a dissipative "adsorption locking" mechanism [52] where adsorbate impedes the motion of the end points.

**Slip traces and surface roughness.** With an eye on capacitive processes, as in the present experiment, it has been pointed out that dislocation endpoints moving along the surface of an idealized crystal with planar facets create slip traces that increase the surface area. Reducing $\gamma$ by capacitive charging would thus reduce an energy barrier for plastic deformation, enhancing the deformation rate or reducing the flow stress magnitude [18, 26, 53]. The argument agrees with our theory inasmuch as mechanical work is again balanced against the work required to increase the surface area. Yet, the conclusions differ: the reduction of $|\sigma^{\text{flow}}|$ upon charging is here predicted irrespective of whether the plastic strain is in compression or in tension. The in situ compression tests on NPG do not support the slip-trace argument: $|\sigma^{\text{flow}}|$ increases upon charging and deformation decreases the surface area. The disagreement was noted by Jin and Weissmüller based on first tentative experiments into the potential-dependent flow behavior of NPG near $E_{zc}$ [26]. On that basis, the authors rejected arguments balancing mechanical work against surface tension as apparently not relevant for the potential-dependent flow. Yet, the apparent contradiction is naturally resolved when one realizes that real surfaces are typically rough and exhibit many pre-existing step edges and kink sites. Dislocations moving along the surface may then not only *create* new step edges – as they invariably do on planar terraces – but also *remove* pre-existing ones [14]. The continuum picture of our theory here appears appropriate; it describes the geometry through the radius $r$ and, thereby, through surface mean curvature. Curvature requires edges, hence roughness, thereby connecting to the atomic-scale picture in a statistical sense. Elongation at constant volume increases the curvature and hence the roughness as well as the net surface area, whereas compression has the opposite effect. These considerations imply that experimental investigations of flow-stress potential coupling for nanostructures with *faceted* surfaces – as opposed to the rough ones of the present material – might reveal qualitatively different behavior. Originally rough surfaces of nanoporous gold have been observed to reconstruct and to develop microfacets when the material acts as a catalyst for CO oxidation [54]. In-situ mechanical tests under controlled gas atmosphere might thus probe a possible distinction between the strength of nanostructures with rough or faceted surfaces in future studies.

**Conclusions.** As advertized in the Introduction, ex-

tremely small nanowires or the very small ligaments of some NPG studies can experience spontaneous irreversible contraction even when there is no external load. Plastic yielding prompted by surface stress has been invoked to explain the observation. Yet, our study does not support surface-stress induced yielding. Instead, since the energies of the initial and final states are governed by the surface tension, it appears appropriate to identify surface tension as the driving force. This in itself does not explain the microscopic mechanism of the spontaneous yielding, since the stresses in the solid are not governed by $\gamma$. Further studies of the issue would seem to be of high interest.

In summary, our experiment provides compelling support of the theory suggesting substantial effects of surface tension on plastic flow at the nanoscale, while rejecting significant contributions by surface stress. This suggests that the impact of capillarity on the size-dependent yield strength and the tension-compression asymmetry of small structures might need to be reconsidered.

[1] R. Shuttleworth. The surface tension of solids. *Proc. Phys. Soc. London A*, 63:444–457, 1950.
[2] J. Weissmüller and J. W. Cahn. Mean stresses in microstructures due to interface stresses: A generalization of a capillary equation for solids. *Acta Mater.*, 45:1899–1906, 1997.
[3] W. Haiss. Surface stress of clean and adsorbate-covered solids. *Rep. Prog. Phys.*, 64:591–648, 2001.
[4] J. K. Diao, K. Gall, and M. L. Dunn. Surface-stress-induced phase transformation in metal nanowires. *Nature Mater.*, 2:656–660, 2003.
[5] W. Liang, M. Zhou, and F. Ke. Shape memory effect in cu nanowires. *Nano Lett.*, 5:2039–2043, 2005.
[6] J. Diao, K. Gall, M.L. Dunn, and J.A. Zimmerman. Atomistic simulations of the yielding of gold nanowires. *Acta Mater.*, 54:643–653, 2006.
[7] D. Farkas, A. Caro, E. Bringa, and D. Crowson. Mechanical response of nanoporous gold. *Acta Mater.*, 61:3249–3256, 2013.
[8] B.-N. D. Ngô, A. Stukowski, N. Mameka, J. Markmann, K. Albe, and J. Weissmüller. Anomalous compliance and early yielding of nanoporous gold. *Acta Mater.*, 93:144–155, 2015.
[9] S. Parida, D. Kramer, C. A. Volkert, H. Rösner, J. Erlebacher, and J. Weissmüller. Volume change during the formation of nanoporous gold by dealloying. *Phys. Rev. Lett.*, 97:035504, 2006.
[10] H.-J. Jin, X.-L. Wang, S. Parida, K. Wang, M. Seo, and J. Weissmüller. Nanoporous Au-Pt alloys as large strain electrochemical actuators. *Nano Lett.*, 10:187–194, 2010.
[11] J. Diao, K. Gall, and M.L. Dunn. Yield strength asymmetry in metal nanowires. *Nano Lett.*, 4:1863–1867, 2004.
[12] W. Zhang, T. Wang, and X. Chen. Effect of surface stress on the asymmetric yield strength of nanowires. *J. Appl. Phys.*, 103, 2008.
[13] C.R. Weinberger and W. Cai. Plasticity of metal nanowires. *J. Mater. Chem.*, 22:3277–3292, 2012.
[14] Xing-Long Ye and Hai-Jun Jin. Electrochemical control of creep in nanoporous gold. *Applied Physics Letters*, 103:201912, 2013.
[15] H. Udin, A.J. Schaler, and J. Wulff. The surface tension of solid copper. *Trans. AIME*, 185:186–190, 1949.
[16] F. H. Buttner, H. Udin, and J. Wulff. The surface tension of solid gold. *Trans. AIME*, 191:1209–1211, 1951.
[17] D. Josell and F. Spaepen. Determination of the interfacial tension by zero creep experiments on multilayers – ii. Experiment. *Acta Metall. Mater.*, 41:3017–3027, 1993.
[18] A. Pfützenreuter and G. Masing. Zunahme der Geschwindigkeit des Plastischen Fliessens von Metallen im Elektrolyten bei der Electrochemischen Polarisation. *Z. Metallkunde*, 51:361–370, 1951.
[19] C. J. van der Wekken. The effect of cathodic polarization on the creep rate of gold. *J. Electrochem. Soc.*, 129:706–711, 1982.
[20] R.M. Latanision, H. Opperhauser Jr., and A.R.C. Westwood. The influence of surface charge density on the fracture of zinc single crystal electrodes. *Scripta Metall.*, 12:475–479, 1978.
[21] N. Mameka, K. Wang, J. Markmann, E. T. Lilleodden, and J. Weissmüller. *Mater. Res. Lett.*, 4:27–36, 2016.
[22] N. Badwe, X. Chen, and K. Sieradzki. Mechanical properties of nanoporous gold in tension. *Acta Materialia*, 129:251–258, 2017.
[23] J. Biener, A. M. Hodge, J. R. Hayes, C. A. Volkert, L. A. Zepeda-Ruiz, A. V. Hamza, and F. F. Abraham. Size effects on the mechanical behavior of nanoporous Au. *Nano Lett.*, 6:2379–2382, 2006.
[24] C. A. Volkert and E. T. Lilleodden. Size effects in the deformation of sub-micron Au columns. *Phil. Mag.*, 86:5567–5579, 2006.
[25] J. Biener, A. V. Hamza, and A. M. Hodge. Deformation behavior of nanoporous metals. In F. Yang and J. C. M. Li, editors, *Micro and Nano Mechanical Testing of Mater. and Devices*, chapter 6, pages 118–135. Springer US, 2008.
[26] H.-J. Jin and J. Weissmüller. A material with electrically tunable strength and flow stress. *Science*, 332:1179–1182, 2011.
[27] N. Mameka, J. Markmann, H.-J. Jin, and J. Weissmüller. Electrical stiffness modulation – confirming the impact of surface excess elasticity on the mechanics of nanomaterials. *Acta Mater.*, 76:272–280, 2014.
[28] S. Sun, X. Chen, N. Badwe, and K. Sieradzki. Potential-dependent dynamic fracture of nanoporous gold. *Nature Mater.*, 14:894–898, 2015.
[29] J. Weissmüller. Electrocapillarity of solids and its impact on heterogeneous catalysis. In R.C. Alkire, L. Kibler, D.M. Kolb, and J. Lipkowski, editors, *Electrocatalysis: Theoretical Foundations and Model Experiments*, Advances in Electrochemical Science and Engineering, pages 163–219. Wiley VCH, Weinheim, Germany, 2013.
[30] Note1. Eqs 4 follow from Eq 20 of Ref [2] when *i*) setting–in line with the present assumption of isotropic surface stress–the axial and radial surface stress components as equal and *ii*) correcting the obvious (surface and bulk stresses need to cancel, which is the central theme in the reference) omission of a minus sign in front of the stress.
[31] J. Weissmüller, H. L. Duan, and D. Farkas. Deformation

of solids with nanoscale pores by the action of capillary forces. *Acta Mater.*, 58:1–13, 2010.

[32] J. W. Cahn and F. Larché. Surface stress and the chemical-equilibrium of small crystals – II. Solid particles embedded in a solid matrix. *Acta Metall.*, 30:51–56, 1982.

[33] J. Weissmüller. Comment on "Lattice contraction and surface stress of fcc nanocrystals". *J. Phys. Chem. B*, 106:889–890, 2002.

[34] Note2. Take the example of a gold nanowire with 111-type surfaces, where roughly $\gamma$ = 1.1 N/m, $f$ = 3.3 N/m [43], $Y$ = 80 GPa and $\nu$ = 0.42 [55]. Even for a nanowire with $r$ as small as 1 nm, the second term in the bracket of Eq 7 is only 0.12 N/m, leaving the first term, $\gamma$, as the dominant contribution. Samples in the experimental part of our work have $r \sim 20$ nm, so that the impact of relaxation on $\gamma$ may be safely neglected.

[35] Note3. When a dislocation line segment d**l** is displaced in its glide-plane by the vector $\delta$**s**, then the work done against the local acting stress **S**$^{\text{net}}$ is $\delta W$ = (**S**$^{\text{net}}\cdot$**n**$\delta A)\cdot$**b**, where **b** denotes the Burgers vector and **n**$\delta A$ = d**l**$\times\delta$**s** with $\delta A$ the area covered [36]. Therefore, when a glide-plane of finite area is swept by the dislocation, the net work scales with the area-integral of **S**$\cdot$**n**, the traction.

[36] P.M. Anderson, J.P. Hirth, and J. Lothe. *"Theory of Dislocations"*, chapter 4.6 and 20.5. Cambridge University Press, New York, 3rd edition, 2017.

[37] S. H. Oh, M. Legros, D. Kiener, and G. Dehm. In situ observation of dislocation nucleation and escape in a submicrometre aluminium single crystal. *Nature Materials*, 8:95–100, 2009.

[38] H.-J. Jin, L. Kurmanaeva, J. Schmauch, H. Rösner, Y. Ivanisenko, and J. Weissmüller. Deforming nanoporous metal: Role of lattice coherency. *Acta Mater.*, 57:2665–2672, 2009.

[39] R. Dou and B. Derby. Deformation mechanisms in gold nanowires and nanoporous gold. *Philosophical Magazine*, 91:1070–1083, 2011.

[40] N. Huber, R. N. Viswanath, N. Mameka, J. Markmann, and J. Weimller. Scaling laws of nanoporous metals under uniaxial compression. *Acta Materialia*, 67:252–265, 2014.

[41] E. Rabkin, H. S. Nam, and D. J. Srolovitz. Atomistic simulation of the deformation of gold nanopillars. *Acta Materialia*, 55:2085–2099, 2007.

[42] P.S. Germain, W.G. Pell, and B.E. Conway. Evaluation and origins of the difference between double-layer capacitance behaviour at Au-metal and oxidized Au surfaces. *Electrochim. Acta*, 49:1775–1788, 2004.

[43] Y. Umeno, C. Elsässer, B. Meyer, P. Gumbsch, M. Nothacker, J. Weissmüller, and F. Evers. Ab initio study of surface stress response to charging. *EPL*, 78:13001, 2007.

[44] M. Smetanin, R. N. Viswanath, D. Kramer, D. Beckmann, T. Koch, L. A. Kibler, D. M. Kolb, and J. Weissmüller. Surface stress-charge response of a (111)-textured gold electrode under conditions of weak ion adsorption. *Langmuir*, 24:8561–8567, 2008.

[45] H. Angerstein-Kozlowska, B.E. Conway, A. Hamelin, and L. Stoicoviciu. Elementary steps of electrochemical oxidation of single-crystal planes of Au part II. A chemical and structural basis of oxidation of the (111) plane. *J. Electroanal. Chem.*, 228:429–453, 1987.

[46] D.M. Kolb and J. Schneider. Surface reconstruction in electrochemistry: Au(100-(5 x 20), Au(111)-(1 x 23) and Au(110)-(1 x 2). *Electrochim. Acta*, 31:929–936, 1986.

[47] R. Holze. Electrochemical thermodynamics and kinetics. chapter 3 Electrode potentials of zero charge. Springer Berlin Heidelberg, 2007.

[48] D.D. Bodé Jr., T. N. Adersen, and H. Eyring. Anion and pH effects on the potentials of zero charge of gold and silver electrodes. *J. Phys. Chem.*, 71:792–797, 1967.

[49] J. Chen, L. Nie, and S. Yao. A new method for rapid determination of the potential of zero charge for gold—solution interfaces. *J. Electroanal. Chem.*, 414:53–59, 1996.

[50] A. Hamelin. The surface state and the potential of zero charge of gold (100): A further assessment. *J. Electroanal. Chem.*, 386:1–10, 1995.

[51] E.D. Shchukin, L.A. Kochanova, and V.I. Savenko. On the mechanism of environment-induced plasticizing under contact interactions. In R.M. Latanision and R.J. Courtel, editors, *Advances in the mechanics and physics of surfaces*, volume 1, pages 111–152. Harwood Academic Publishing, 1981.

[52] A. R. C. Westwood. The Rebinder effect and the adsorption-locking of dislocations in lithium fluoride. *Phil. Mag.*, 7:633–649, 1962.

[53] R.M. Latanision, H. Opperhauser Jr., and A.R.C. Westwood. Electrocapillarity and the microhardness of zinc monocrystal electrodes. In *Proceedings of the 5th International Congress on Metallic Corrosion, NACE, Houston, TX*, pages 111–114, 1974.

[54] T. Fujita, P. Guan, K. McKenna, X. Lang, A. Hirata, L. Zhang, T. Tokunaga, S. Arai, Y. Yamamoto, N. Tanaka, Y. Ishikawa, N. Asao, Y. Yamamoto, J. Erlebacher, and M.W. Chen. Atomic origins of the high catalytic activity of nanoporous gold. *Nature Materials*, 11:775–780, 2012.

[55] W.F. Gale and T.C. Totemeier, editors. *Smithells Metals Reference Book*. Elsevier Butterworth-Heinemann, 8th edition, 2004.

[56] H. Tobias and A. Soffer. The immersion potential of high surface electrodes. *J. Electroanal. Chem.*, 148:221–232, 1983.

[57] U.W. Hamm, D. Kramer, R.S. Zhai, and D.M. Kolb. The pzc of Au(111) and Pt(111) in a perchloric acid solution: An ex situ approach to the immersion technique. *J. Electroanal. Chem.*, 414:85–89, 1996.

[58] Y. Xue, J. Markmann, H.L. Duan, J. Weissmüller, and P. Huber. Switchable imbibition in nanoporous gold. *Nature Comm.*, 5:4237, 2014.

[59] L. Lührs, C. Soyarslan, J. Markmann, S. Bargmann, and J. Weissmüller. Elastic and plastic poissons ratios of nanoporous gold. *Scripta Mater.*, 110:65–69, 2016.

## ACKNOWLEDGEMENTS

This work was supported by Deutsche Forschungsgemeinschaft through SFB 986 "Taylor-Made Multiscale Materials Systems - M$^3$", subprojects B2 and B8.

## AUTHOR CONTRIBUTIONS

N.M. designed and performed the experiments, analyzed the results, and wrote the manuscript. J.M. designed the testing protocols, J.W. the theory. All authors discussed results and manuscript.

## METHODS

**Preparation of NPG by dealloying.** Master alloys $Au_{25}Ag_{75}$ were prepared according to Ref. [26], except that wire drawing and sectioning by a wire saw were used to make cylindrical samples, 1.17 – 1.37 mm in diameter and 1.90 – 2.10 mm in length. Electrochemical dealloying in 1 M HClO4 (60% HClO4, ACS grade, Merck) at ambient temperature used a potential of 1.26 V vs. SHE. Reference and counter electrodes (RE and CE) for dealloying were pseudo Ag/AgCl (+0.515 vs. SHE) and a coiled Ag wire (99.9985%, Alfa Aesar), respectively. Using 1 M $HClO_4$ prepared from high purity $HClO_4$ (70%, Suprapur, Merck), the as-dealloyed samples were reduced during 15 potential cycles between 0.01 and 1.01 V with a scan rate of 5 mV/s; this lead to mean ligament diameter $L \sim 20$ nm. Cycling 15 times between 0.01 and 1.51 V at 5 mV/s led to $L \sim 35-45$ nm. Rinsing with ultrapure water (Ultra Clear TWF UV TM, Siemens) and drying for > 2 days in Ar (5.0) flow at room temperature followed. Samples with $L \sim 150$ nm were prepared by annealing $L = 20$ nm NPG at 400°C for 1 h in Ar. Solid fractions, as determined from external sample dimensions and mass, were $\varphi = 0.27 \pm 0.01, 0.29 \pm 0.02, 0.32 \pm 0.03$ for samples with $L = 20, 40, 150$ nm, respectively. Mean ligament diameters, $L$, were estimated from scanning electron micrographs. All potentials in this work are referred to the standard hydrogen electrode (SHE).

**Immersion method.** As an approach to the potential of zero charge, $E_{zc}$, of NPG under experimental conditions we exploited the large surface area and used two variants of the immersion method [56, 57]. Electrochemically reduced and dried NPG with $L = 40$ nm was connected as the WE of a three-electrode cell, initially without contact to the electrolyte, and was then rapidly immersed (for wetting kinetics see Ref [58]). In the "open-circuit immersion" method, zero-current galvanostatic control was maintained, and the initial value of the potential transient supplied $E_{zc}$. The "charge integration" method studies immersion under potentiostatic control at a series of fixed immersion potentials, $E_{im}$. $E_{zc}$ was determined by linear regression using the net charges, $Q(E_{im})$, as determined by integration of the current transient. Fig 6 b shows that the two methods are highly consistent, for example the charge integration suggests $E_{zc} = 0.77 \pm 0.19$ V (errors derived from linear regression) for $HClO_4$, as compared to $0.77 \pm 0.10$ V (errors derived from scatter for repeated experiments) from open-circuit immersion.

**In situ mechanical testing.** Compression tests at ambient temperature used a Zwick Z010 TN frame, at constant engineering strain rate $10^{-5}$ s$^{-1}$. The strain was measured by a laser extensometer. The in situ setup (Fig 3) used a glass cuvette filled with electrolyte with a cold-worked Pt plate as lower load surface and electrical contact. The load was applied via a glass rod. A gold wire wrapped with porous carbon cloth formed the CE, Ag/AgCl (Dri-Ref, World Precision Instruments) served as the RE, and a potentiostat (PGSTAT 302N, Metrohm) controlled the potential. Throughout this work we specify engineering strains and stresses. Since cross-sections vary little [38, 59], engineering and true stresses agree closely.

# Supplementary Information for "On the impact of capillarity for strength at the nanoscale"

Nadiia Mameka[1*], Jürgen Markmann[1,2], and Jörg Weissmüller[1,2*]
1–Institute of Materials Research, Materials Mechanics,
Helmholtz-Zentrum Geesthacht, Geesthacht, Germany and
2–Institute of Materials Physics and Technology,
Hamburg University of Technology, Hamburg, Germany*

## I. CYCLIC VOLTAMMETRY OF NPG IN DIFFERENT ELECTROLYTES

Figure S1 shows cyclic voltammograms of nanoporous gold (NPG) in the electrolytes of our study. This data was used for defining the interval of nominally capacitive charging, for use in the in situ compression tests.

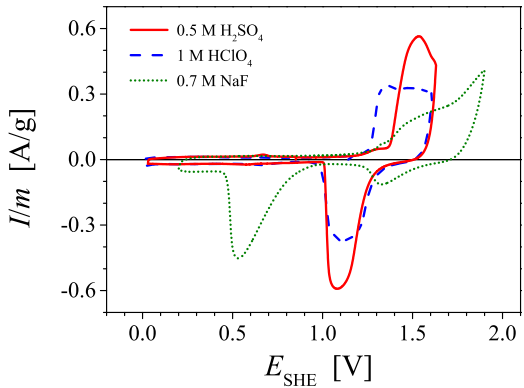

FIG. S1. Cyclic voltammograms of NPG with mean ligament diameter of 40 nm in the studied electrolytes of $H_2SO_4$, $HClO_4$, and NaF. Scan rate 5 mV/s. Current $I$ is normalized to a sample mass $m$ in each single experiment.

## II. IN SITU COMPRESSION TESTS DURING CAPACITIVE PROCESSES

On top of the in situ compression tests shown in the main text, additional tests on NPG samples were performed in smaller potential intervals, excluding the region of oxygen electrosorption. The purpose of these experiments was to verify that *i.)*, the sign inversion of the flow-stress potential response is recovered when the potential step sequence is inverted and that *ii.)*, consistent results for the response are also obtained in the initial stages of plastic deformation, at small strain. Figure S2 illustrates the results of an experiment with mean ligament size $L$ = 40 nm in 1 M $HClO_4$ and of another experiment with $L$ = 30 nm in 0.5 M $H_2SO_4$. Parts a) and b) of the figure show the stress-strain curves (red graphs) along with the potential step protocols (blue graphs). In both compression tests the potential scan started with a hold at the potential of zero charge, $E_{zc}$, for NPG in the respective electrolyte. The potential, $E$, was then decreased to a more negative value and subsequently stepped upward (anodic) to an upper vertex of 1.0 V. In Fig S2a the scan direction was then inverted, so that the potential step series continued negative-going (cathodic). In Fig S2b the potential cycles were started at lesser strain, at about 4 % of engineering strain.

As can be seen from Fig S2c, the trends for the flow-stress potential coupling parameter $\delta\sigma/\delta E$ versus $E$ are indeed reproducible on the anodic and cathodic step series: Consistent with the graph for 1 M $HClO_4$ in Fig 6a of the main text, the coupling approximates zero when $E_{zc}$ is approached from below, and the behavior is consistent during the backwards scan. The numerical values of the coupling during anodic and cathodic scan differ. This is consistent with the flow stress increase during ongoing compression and with our observation, see the main text, that the coupling scales with the net value of the flow stress. In fact, the two branches coincide in the plot of the normalized response parameter, $\sigma^{-1}\delta\sigma/\delta E$, in Fig S2d.

Fig S2b shows that the findings from the main text apply also when the potential jumps are started at an earlier stage of deformation, namely already at 4% strain. Again, the flow-stress potential response – here in 0.5 M $H_2SO_4$ – is negative for $E < E_{zc}$ and positive for $E > E_{zc}$.

A comparison of the absolute values of the normalized response parameter with those of Fig 6a in the main text (grey symbols in Fig S2d) shows excellent agreement.

As in the main text, we also compared the experimental coupling parameters to the theory of Eq 12, which is represented by the solid lines in Fig S2c. For the experiment of Fig S2b the response near the pzc is well compatible with the theory. The data derived from the experiment of Fig S2a shows a smaller-than-predicted $\delta\sigma/\delta E$ during the anodic potential scans, whereas the later, cathodic series gives a stronger response than predicted. In view of the good agreement of the normalized parameters with those in Fig 6a we conclude that the behavior of these two samples is consistent with the reports in the main text, except for a somewhat lesser or greater flow stress.

* Authors for correspondence: nadiia.mameka@hzg.de, weissmueller@tuhh.de

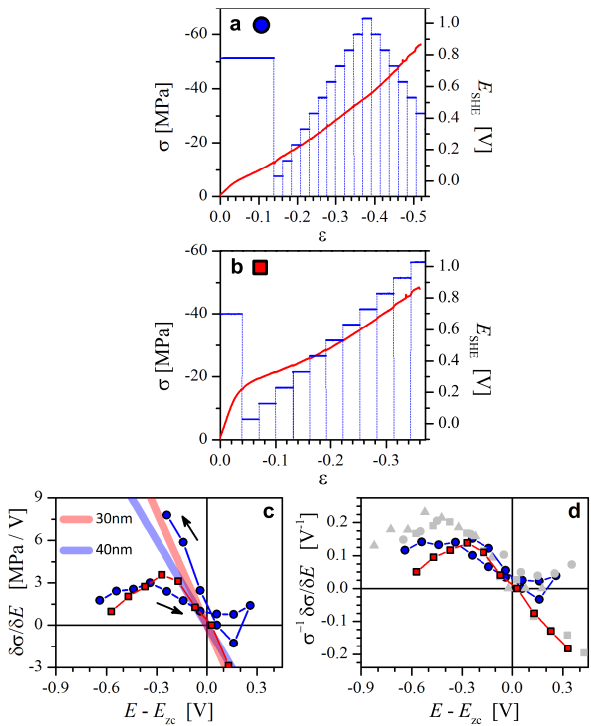

FIG. S2. In situ compression tests for NPG at constant engineering strain rate of $10^{-5}$ s$^{-1}$ but with different potential step protocols. **a**) Test for a sample with ligament size $L = 40$ nm in 1 M HClO$_4$, with potential steps exploring anodic (positive-going) and reversed cathodic directions. Red: graph of stress $\sigma$ versus strain $\varepsilon$; blue: electrode potential $E$ versus the standard hydrogen electrode (SHE). **b**) Test as in a), but for $L = 30$ nm and in 0.5 M H$_2$SO$_4$, here with potential jumps (in anodic direction) starting already at strain 4%. **c**) Flow stress-potential response, $\delta\sigma/\delta E$, versus the potential, $E - E_{zc}$, relative to the potential of zero charge, from the experiments in a) (blue circles) and in b) (red squares). Arrows show directions of the potential steps. Bold lines: predicted coupling strength near $E_{zc}$ from Eq (12) of the main text, using the capacitance value $c = 40$ $\mu$F/cm$^2$ and ligament sizes as indicated in legend. **d**) Normalized flow stress response versus potential. Grey symbols: data from the in-situ tests compiled in Fig. 6 of the main text. Note the mutual consistency of all data sets, irrespective of ligament size and electrolyte.

## III. SURFACE AREA VARIATION DURING COMPRESSION

Figure S3a displays the surface area, $A$, that was evaluated in situ during compression in 1 M HClO$_4$. The experimental procedure was as follows: We used the capacitance ratio method [1], which determines $A$ from $A = C/c_{DL}$ with $C$ the net capacity and $c_{DL}$ the specific double-layer capacitance, $c_{DL} = 40$ $\mu$F/cm$^2$ [2]. Electrochemical impedance spectrometry (EIS) in the frequency range 0.1 – 1 Hz and at the DC potential 0.8 V provided $C$ via the procedures from Ref [3]. The impedance spectra were run continuously during the whole deformation experiment and the DC potential of 0.8 V was shortly (60 – 80 s) imposed on the sample during recording of each single EIS data point. The values of $A$ were then normalized by the solid volume (sample mass divided by mass density of gold) of the NPG samples.

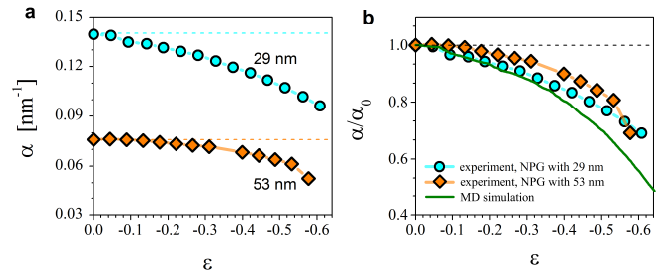

FIG. S3. **a**, Evolution of volume-specific surface area (per solid volume), $\alpha$, of NPG with compressive strain $\varepsilon$ for two different ligament sizes as measured by electrochemical impedance spectroscopy technique in 1 M HClO$_4$. **b**, Comparison with changes in $\alpha$ of "dry" NPG upon compression as obtained by molecular dynamics simulations (MD) in Ref. [4] (green solid line). Both experimental and MD values have been scaled to the starting value of $\alpha$ before compression, $\alpha_0$. Dashed lines emphasize initial value, solid lines are guides to the eye. Note the excellent agreement between experiment and MD results.

The graphs of $A$ versus the compressive strain in Fig S3a show a consistent reduction of area during plastic deformation, consistent with a key assumption in our theory. The graph of relative change in surface area during compression, Fig S3b, affords a verification of the consistency. Superimposed to the data is here a graph representing the molecular dynamics simulation (MD) of plastic deformation of NPG during compression in Ref [4] (Fig. 5b there). It is seen that the graphs from the experiment and simulation are in good agreement, specifically in the early stages of compression. Note that the area may be affected by the formation of new contacts between ligaments through cold welding. Yet, this process is expected to prevail only at the later stages of compression. The variation in surface area is therefore significant as an indication of thickening of the ligaments during compression, see our theory.

It is known that exposure to electrolyte and potential can leads to coarsening of the ligaments of NPG [5, 6]. Changes in microstructure before and after applying the potential steps of the experiment in Fig S2 have been explored in the following way: An as-prepared sample was cleaved and one of the two halves was exposed to potential cycles as in an in situ compression test, whereas the other saw no contact with electrolyte. Both halves were then investigated side-by-side in a scanning electron microscope and the identical image analysis was used to quantify a distribution of sizes. Figures S4a and b shows the corresponding scanning electron micrographs (SEMs). No change in ligament diameter is apparent to the naked eye.

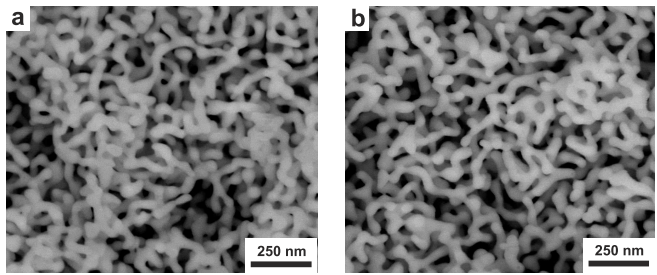

FIG. S4. Scanning electron micrographs of microstructure before (**a**) and after (**b**) applying potential jumps representative of a mechanical in-situ tests. Potential-time protocol in 1 M $HClO_4$ was identical with Fig S2 and with Fig 5 of the main text, except that the series here ended after 1.0 V was reached. This replicates the variation in the capacitive regime, which underlies Fig 6a in the main text and the discussion of capacitive processes there. Parts a) and b) image two opposing faces of a cleavage surface of the same sample, treated differently. Analysis of the mean ligament size in the two images supplied the values 41.7 and 43.1 nm, respectively.

For a precise assessment of possible coarsening, we performed a stereologic analysis of the images by the image processing software ImageJ [7]. SEMs of larger areas, containing roughly 1000 ligaments, were binarized using identical threshold values and the resulting images were then analyzed by ImageJ's BoneJ Thickness plugin [8–10]. In its implementation for 2D images, the algorithm uses the method of Ref [11], computing a local size measure as the diameter of the largest inscribed circle at any point and specifying the mean size as the average of the diameter distribution.

Analysis of the mean ligament size in the images of Fig S4a and b supplied the values 41.7 and 43.1 nm, respectively. In view of the microstructural heterogeneity that is present even in the quite well-defined structure of NPG, the two numbers may be considered as consistent with little, if any, coarsening. The relative change in ligament size is certainly smaller than the relative change in specific surface area of Fig S3. This implies minimal or no effect of the applied potentials on the porous network in our study. The observed decreasing of the surface area may thus be dominantly attributed to the compressive strain.

## IV. LINK BETWEEN FLOW STRESS AND YIELD STRENGTH IN NPG

A connection between flow stress and yield strength of NPG is very well illustrated by a compressive stress-strain diagram with unload-reload cycles (Fig. S5). In the compression test, a dry sample with ligament diameter of about 40 nm was continuously loaded with engineering strain rate of $10^{-4}$ $s^{-1}$ and subjected to intermediate load/unload segments. During reloading, a strain level was incrementally increased followed by unloading up to a minimum stress value of ~ 2 MPa. The further details of the experiment can be found in Ref [12]. Upon reloading the NPG specimen suffered plastic yielding at the flow stress value that was reached just before the unload. This implies that the flow stress at a certain state of plastic strain represents the yield strength of the material in that strain state. Thereby, it emphasis a strong coupling between strength and plastic flow in NPG.

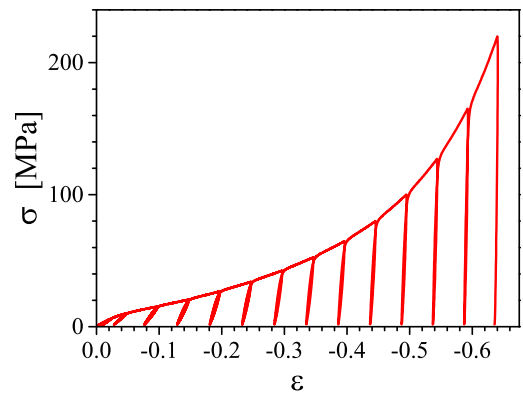

FIG. S5. Load-unload stress-strain data of dry NPG with $L$ = 40 nm in compression. Engineering strain rate $10^{-4}$ $s^{-1}$. Note the material's yielding at the beginning of reload segment that coincide with the last flow stress value prior to the unload.

[1] S. Trasatti and O.A. Petrii. Real surface area measurements in electrochemistry. *J. Electroanal. Chem.*, 327 (1-2):353–376, 1992.
[2] P.S. Germain, W.G. Pell, and B.E. Conway. Evaluation and origins of the difference between double-layer capacitance behaviour at Au-metal and oxidized Au surfaces. *Electrochim. Acta*, 49(11):1775–1788, 2004.
[3] E. Rouya, S. Cattarin, M. L. Reed, R. G. Kelly, and G. Zangari. Electrochemical characterization of the surface area of nanoporous gold films. *J. Electrochem. Soc.*, 159(4):K97–K102, 2012.
[4] B.-N. D. Ngô, A. Stukowski, N. Mameka, J. Markmann, K. Albe, and J. Weissmüller. Anomalous compliance and early yielding of nanoporous gold. *Acta Mater.*, 93:144–155, 2015.
[5] Y. Ding, Y.-J. Kim, and J. Erlebacher. Nanoporous gold leaf: "Ancient technology"/Advanced material. *Advanced Materials*, 16(21):1897–1900, 2004.

[6] A. Sharma, J.K. Bhattarai, A.J. Alla, A.V. Demchenko, and K.J. Stine. Electrochemical annealing of nanoporous gold by application of cyclic potential sweeps. *Nanotechnology*, 26(8):085602, 2015.

[7] C.A. Schneider, W.S. Rasband, and K.W. Eliceiri. NIH Image to ImageJ: 25 years of image analysis. *Nature Methods*, 9(7):671–675, 2012.

[8] M. Doube, M.M. Klosowski, I. Arganda-Carreras, F.P. Cordeliéres, R.P. Dougherty, J.S. Jackson, B. Schmid, J.R. Hutchinson, and S.J. Shefelbine. BoneJ: Free and extensible bone image analysis in ImageJ. *Bone*, 47(6): 1076–1079, 2010.

[9] R. Dougherty and K. H. Kunzelmann. Computing local thickness of 3D structures with ImageJ. *Microscopy and Microanalysis*, 13(S02):1678–1679, 2007.

[10] T. Hildebrand and P. Rüegsegger. A new method for the model-independent assessment of thickness in three-dimensional images. *Journal of Microscopy*, 185(1):67–75, 1997.

[11] N.J. Garrahan, R.W.E. Mellish, S. Vedi, and J.E. Compston. Measurement of mean trabecular plate thickness by a new computerized method. *Bone*, 8(4):227–230, 1987.

[12] N. Huber, R. N. Viswanath, N. Mameka, J. Markmann, and J. Weißmüller. Scaling laws of nanoporous metals under uniaxial compression. *Acta Mater.*, 67(0):252–265, 2014.